\documentclass[final,5p,times,twocolumn]{elsarticle}

\usepackage{amssymb}
\usepackage{amsthm}
\usepackage{epstopdf}
\usepackage{amsmath}
\usepackage{subcaption}
\usepackage{comment}
\newtheorem{theorem}{Theorem}[section]

\newtheorem{remark}[theorem]{Remark}
\newtheorem{definition}[theorem]{Definition}

\begin{document}

\begin{frontmatter}



\title{Data-Enabled Predictive Control for Fast Charging of Lithium-Ion Batteries with Constraint Handling}
%\tnoteref{t1}}
%\tnotetext[t1]{This paper has not been submitted to any conferences or journals previously.}

\author[affi1]{Kaixiang Zhang\corref{cor1}}
  \ead{zhangk64@msu.edu}
\author[affi1]{Kaian Chen\corref{cor1}}
  \ead{chenkaia@msu.edu}  
\author[affi2]{Xinfan Lin}
  \ead{lxflin@ucdavis.edu}
\author[affi3]{Yusheng Zheng}
  \ead{zhengyusheng@cqu.edu.cn}
\author[affi4]{Xunyun Yin}
  \ead{Xunyuan.Yin@ntu.edu.sg}
\author[affi3]{Xiaosong Hu}
  \ead{xiaosonghu@ieee.org}
\author[affi5]{Ziyou Song}
  \ead{ziyou@nus.edu.sg}
\author[affi1]{Zhaojian Li\corref{cor2}}
  \ead{lizhaoj1@msu.edu}
\cortext[cor1]{The authors made equal contributions to this paper.}
\cortext[cor2]{Zhaojian Li is the corresponding author.}

\address[affi1]{Department of Mechanical Engineering, Michigan State University, East Lansing, 48824, MI, USA.}

\address[affi2]{Department of Mechanical and Aerospace Engineering, University of California Davis, Davis, 95616, CA, USA.}

\address[affi3]{Department of Mechanical and Vehicle Engineering, Chongqing University, Chongqing, 400044, China}

\address[affi4]{School of Chemical and Biomedical Engineering, Nanyang Technological University, 637459, Singapore}

\address[affi5]{Department of Mechanical Engineering, National University of Singapore, 117575, Singapore}

%\affiliation[affi1]{organization={The Department of Mechanical Engineering, Michigan State University},
%            city={East Lansing},
%            postcode={48824},
%            state={MI},
%            country={USA}}

%\affiliation[affi2]{organization={The Department of Mechanical and Aerospace Engineering, University of California Davis},
%            city={Davis},
%            postcode={95616},
%            state={CA},
%            country={USA}}

%\affiliation[affi3]{organization={The Department of Mechanical and Vehicle Engineering, Chongqing University},%Department and Organization
%            city={Chongqing},
%            postcode={400044}, 
%            country={China}}

%\affiliation[affi4]{organization={The School of Chemical and Biomedical Engineering, Nanyang Technological University},%Department and Organization
%            postcode={637459}, 
%            country={Singapore}}

%\affiliation[affi5]{organization={The Department of Mechanical Engineering, National University of Singapore},%Department and Organization
%            postcode={117575}, 
 %           country={Singapore}}

\begin{abstract}
Fast charging of lithium-ion batteries has gained extensive research interest, but most of the existing methods are either based on simple rule-based charging profiles or require explicit battery models that are non-trivial to identify accurately. In this paper, instead of relying on parametric battery models that are costly to derive and calibrate, we employ a novel data-enabled predictive control (DeePC) paradigm to perform safe and optimal fast charging for lithium-ion batteries. The developed DeePC methodology is based on behavioral system theory and directly utilizes the input-output measurements from the battery system to predict the future trajectory and compute the optimal control policy. Constraints on input current and battery states are incorporated in the DeePC formulation to ensure battery fast charging with safe operations. Furthermore, we propose a principal component analysis based scheme to reduce the dimension of the optimization variables in the DeePC algorithm, which significantly enhances the computation efficiency without compromising the charging performance. 
%Numerical simulations on a high-fidelity battery simulator validate the efficacy of the proposed fast charging strategy as compared to the widely adopted constant-current constant-voltage charging protocols.
Numerical simulations are performed on a high-fidelity battery simulator to validate the efficacy of the proposed fast charging strategy.
\end{abstract}




\end{frontmatter}


\section{Introduction}
\label{sec:intro}
Due to its zero-emission potential and high energy efficiency \cite{Battery,WassiliadisJES2021,XieJES2020}, lithium-ion (Li-ion) batteries have taken the forefront as the power source for electric vehicles (EVs). However, compared to gasoline-powered vehicles that can be fully refueled in minutes, the long charging time has been a major hurdle to the wider public adoption of Li-ion battery-powered EVs. Conventionally, the charging time of Li-ion battery can be reduced by increasing the charging current rate \cite{Alg4BMS}, which will, however, accelerate the battery degradation, shorten its lifespan, and bring potential safety hazards. Therefore, the development of fast charging technologies for Li-ion batteries needs not only to address the charging time  but also to maintain the battery in a nondestructive charging mode for an extended lifespan.

Currently, constant-current constant-voltage (CC-CV) \cite{CCCV} charging is the most commonly used protocol for Li-ion batteries in industry. It consists of two charging phases: a \textit{constant-current} phase in which the battery voltage is raised to a predetermined value,  followed by a \textit{constant-voltage} phase in which the voltage remains constant until the current falls below a predetermined threshold. In addition, some variants, such as multistage constant current protocols \cite{MCC1,MCC2} and pulse charging protocols \cite{PulseCharging1,PulseCharging2}, are designed to achieve faster charging and/or less battery degradation. However, it is empirically challenging and time-/cost- intensive to design and calibrate the appropriate charging profiles for these protocols. Moreover, they are unable to explicitly handle system constraints and not optimal with regard to metrics such as charging time, safety, and degradation.

To overcome the aforementioned difficulties, model-based optimization methods have been developed, where the commonly used battery models can be roughly grouped into equivalent circuit models (ECMs) and electrochemical models (EMs). ECMs integrate the models of different  electrical components, including voltage source, resistors, and capacitors. With appropriate parameterization, the ECMs are able to characterize the electrical dynamics and some internal states of batteries, such as state-of-charge (SOC) \cite{equC-SOC1,equC-SOC2}, terminal voltage \cite{equC}, impedance \cite{equC-SOC2}, and heat generation \cite{equC-temp}. Based on the established ECMs, the optimization techniques, such as dynamic programming \cite{optimization-DP}, genetic algorithm \cite{equC-temp,equC-VCP}, fuzzy control \cite{optimization-fuzzy}, and min-max strategy \cite{optimization-minmax}, can be leveraged to generate optimal charging protocols for fast charging, while maintaining battery safety and mitigating degradation. However, ECMs can only mimic the external characteristics such as the voltage response of batteries and they are not able to model the electrochemical processes inside the cell, especially the side reactions triggered by fast charging. Therefore, charging protocols developed through ECMs will not be able to adequately suppress  the side reactions due to the lack of explicit constraints on internal electrochemical states.

Conversely, the EMs \cite{electrochemical-doyle} are developed based on first principles and can capture the internal mechanisms, e.g., ion diffusion and de-/intercalation in the electrodes and ion transport in the electrolyte. Hence, EMs can provide insights into electrochemical states of the battery. However, EMs are complex to establish and the model calibration is also a non-trivial task \cite{Alg4BMS}. In addition, it is challenging to design suitable controllers based on such highly nonlinear and complex EMs. Attempts have been made to reduce EM complexity by the simplification of battery structure \cite{electrochemical-P2D} and internal dynamics \cite{electrochemical-SPM}. With the simplified models, many closed-loop optimal controls are formulated for battery fast charging. For instance, a fast charging protocol was developed in \cite{electrochemical-Segment} based on a simplified EM, where the side reactions are controlled explicitly. By incorporating thermal dynamics into a similar EM, a reference governor based nonlinear model predictive control (MPC) \cite{electrochemical-NMPC} was proposed to reduce charging time and mitigate temperature-dependent degradation caused by lithium plating and the growth of solid electrolyte interphase. Although model-based approaches provide optimal solutions for fast charging in theory, robustness, and optimality can rarely be accomplished in practice because the Li-ion battery dynamics are constantly changing due to distinct operation modes and uncertainties.

The aforementioned observations, along with advancements in computing and sensing technology, have cultivated a trend toward model-free and data-driven control techniques. For example, reinforcement learning~\cite{DRL-BFC} and Neural Network-Based approximation~\cite{NN_BFC} have been utilized to design optimal charging schemes for batteries. Although these techniques can use experimental data to learn a model or simulate uncertain behaviors, they require tedious tuning and training procedures and have limited interpretability. Recently, Data-EnablEd Predictive Control (DeePC) emerged as a promising model-free optimal control paradigm that directly uses input-output data to achieve safe and optimal control of unknown systems \cite{DeepC}. In contrast to traditional model-based control schemes that rely on an accurate parametric model~\cite{MPC_book}, DeePC leverages behavioral system theory \cite{behavior} and Willems' fundamental lemma \cite{Fundamental-Lemma} to implicitly describe the system trajectories using collected input-output data \cite{DeepC}. It has been revealed that DeePC is equivalent to an MPC formulation for linear time-invariant (LTI) systems~\cite{DeepC-Equiv}, and its application to nonlinear and stochastic systems shows promising performance with the aid of different regularization techniques \cite{DeePC-Stochastic,DeePC-Nonlinear}. The DeePC is especially appealing to systems where the model is difficult or expensive to obtain, and it has been successfully implemented in different applications, including unmanned aerial vehicles \cite{DeepC}, power grids \cite{DeePC-PowerGrid}, and connected and automated vehicles \cite{DeePC-CAV}.
However, the high dimension of optimization variables in DeePC (usually higher than those in its MPC counterpart) makes it computationally expensive for real-time implementation. Therefore, it is crucial to develop effective methods to reduce the computational cost, enabling its application and widespread adoption in actual engineering systems.

In this paper, we develop an efficient DeePC approach for fast charging of Li-ion batteries with constraint handling capabilities. Specifically, without relying on an explicit, complex battery model, an optimal charging controller is designed by utilizing measurable battery input and output data. Both input current and battery internal state constraints are explicitly considered in the DeePC formulation, ensuring a safe charging operation. Meanwhile, a novel principal component analysis (PCA) based scheme is developed to reduce the dimension of optimization variables and thus alleviate the computational burden of the control paradigm. 

Our main contributions include the following. First, this paper spearheads the application of the DeePC framework to address the safe and fast charging of Li-ion batteries. The cumbersome modeling and validation processes of ECMs and EMs are circumvented by utilizing offline data collection and processing. Second, we develop a novel PCA-based dimension reduction method for the ease of DeePC online implementation, achieving much improved computation efficiency without degrading the system performance. 
%Last but not least, the efficacy of the proposed scheme on fast charging of Li-ion batteries with safety related constraints is evaluated in a high-fidelity battery simulator, showing that the developed method not only achieves safe and efficient charging but is also lightweight in computation.
Last but not least, the efficacy of the proposed scheme on fast charging of Li-ion batteries with safety related constraints is evaluated in a high-fidelity battery simulator. In comparison to the widely adopted CC-CV charging protocol and the conventional MPC method, the proposed scheme can achieve safer and more efficient charging.

The rest of this paper is organized as follows. Section \ref{sec:background} reviews the DeePC algorithm, as well as its regularized version. Section \ref{sec:DeePC} introduces the DeePC algorithm for optimal charging of Li-ion battery systems. Section \ref{sec:eff-DeePC} presents the PCA-based dimension reduction method for the optimization variables of DeePC. Section \ref{sec:simulation} shows the simulation results, and conclusions are made in Section \ref{sec:conclusions}.

\textbf{Nomenclature:}
We denote $\mathbb{Z}_{+}$ and $\mathbb{R}$ as the set of positive integers and real numbers, respectively. $0$ used in $[\cdot]$ is denoted as a zero vector with compatible size. The weighted 2-norm of a vector $x$ is denoted by $\|x\|_P\triangleq \sqrt{x^{\top}Px}$, where $P \succ (\succcurlyeq) 0$ is a positive (semi-)definite matrix. Given a signal $\omega \in \mathbb{R}^{n}$ and two integers $i, j\in \mathbb{Z}_{+}$ with $i\le j$, we denote by $\omega_{\left[i, j\right]}$ the restriction of $\omega$ to the interval $\left[i, j\right]$, namely, $\omega_{\left[i, j\right]} := \begin{bmatrix}
	\omega^{\top}(i), \omega^{\top}(i+1), \cdots, \omega^{\top}(j)
\end{bmatrix}^{\top}$. To simplify notation, we will also use $\omega_{\left[i, j\right]}$ to denote the sequence $\left\lbrace \omega(i), \cdots, \omega(j)\right\rbrace$. Given a matrix $A\in\mathbb{R}^{n\times m}$, we denote by $A_{[i,j]},1\leq i<j\leq m$ the restriction of $A$ to the interval from $i$th column to $j$th column. We refer to $x(t)$ the system measurement at time instant $t$, while quantity $x_{i|t}$ is a prediction value $i$ time-steps ahead of current time instant $t$. 

\section{Preliminaries}\label{sec:background}

\begin{figure*}[h]
  \centering
  \includegraphics[width=0.85\textwidth]{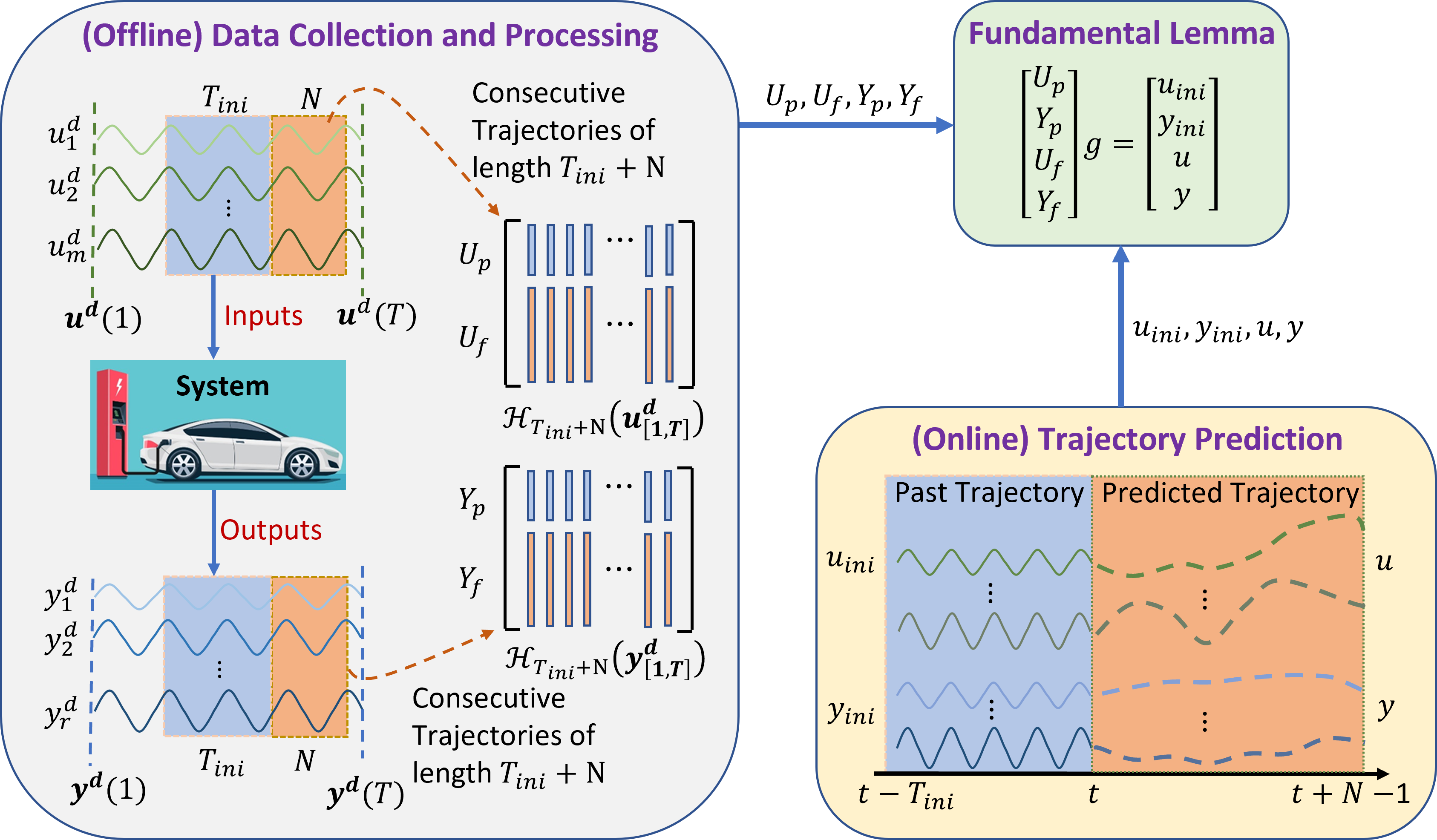}
  \caption{{\small Schematic diagram of the data-enabled predictive control.}}\label{fig:FundamentalLemma}
\end{figure*}
\subsection{Data-EnablEd Predictive Control (DeePC)}
DeePC was originally developed for LTI systems as a model-free optimal control strategy that does not rely on an explicit parametric model. Instead, it leverages Willems' fundamental lemma \cite{Fundamental-Lemma} and generates control inputs (and resultant system outputs) compatible with the pre-collected system data. Specifically, let ${u^d}=[u_1^d,\,u_2^d,\,\cdots, u_m^d]^\text{T}\in\mathbb{R}^m$ be the control inputs and let ${y^d}=[y_1^d,\,y_2^d,\,\cdots, y_r^d]^\text{T}\in\mathbb{R}^r$ be the system outputs. As shown in Figure~\ref{fig:FundamentalLemma}, an input/output data sequence of length $T$ is first collected, and the consecutive trajectories of length $T_\mathrm{ini}+N$ are extracted to form the following Hankel matrices:

\begin{equation}\label{eq:Hankel}
    \begin{aligned}
    \mathcal{H}_{L}({u^d_{[1,T]}})&=\begin{bmatrix}
    {u^d}(1) & {u^d}(2) &\cdots &{u^d}(T-L+1)\\
    {u^d}(2) & {u^d}(3) &\cdots &{u^d}(T-L+2)\\
    \vdots &\vdots &\ddots &\vdots\\
    {u^d}(L) & {u^d}(L+1) &\cdots &{u^d}(T)\\ \end{bmatrix}=\begin{bmatrix}U_{\mathrm{p}}\\{U}_{\mathrm{f}}
    \end{bmatrix}, \\
    \mathcal{H}_{L}({y^d_{[1,T]}})&=\begin{bmatrix}
    {y^d}(1) & {y^d}(2) &\cdots &{y^d}(T-L+1)\\
    {y^d}(2) & {y^d}(3) &\cdots &{y^d}(T-L+2)\\
    \vdots &\vdots &\ddots &\vdots\\
    {y^d}(L) & {y^d}(L+1) &\cdots &{y^d}(T)\\
    \end{bmatrix}=\begin{bmatrix}{Y}_{\mathrm{p}}\\{Y}_{\mathrm{f}}
    \end{bmatrix},
    \end{aligned}
\end{equation}
where $L=T_\mathrm{ini}+N$, and ${U}_{\mathrm{p}}$ denotes the first $T_\mathrm{ini}$ block rows of $\mathcal{H}_{L}(u^d_{[1,T]})$ and represents the ``past'' segment of the input trajectory  whereas ${U}_{\mathrm{f}}$ denotes the last $N$ block rows of $\mathcal{H}_{L}(u^d_{[1,T]})$ and represents the ``future'' segment of the input trajectory. The matrices ${Y}_{\mathrm{p}}$ and ${Y}_{\mathrm{f}}$ are similarly defined. Essentially, the columns of $\mathcal{H}_{L}(u^d_{[1,T]})$ and $\mathcal{H}_{L}(y^d_{[1,T]})$ can be viewed as libraries of input/output trajectories of length $L$, which are further partitioned into segments with length $T_\mathrm{ini}$ (representing ``past'') and length $N$ (representing ``future'') to be compatible with the online optimization formulation which will be discussed in later sections. We next introduce the following concept necessary for establishing our results.

\begin{definition}[Persistently Exciting]\label{def1}
A signal sequence $w_{[1,T]}$ is persistently exciting of order $L$ $(L\leq T)$ if the Hankel matrix $\mathcal{H}_{L}(w_{[1,T]})$ is of full row rank.
\end{definition}

\begin{remark}[Minimal Length of $T$ \cite{DeepC}]\label{remk1}
    In order to make $w_{[1,T]}$ persistently exciting of order $L$, it must have $T\geq (m+1)L-1$, that is, the input signal sequence $w_{[1,T]}$ should be sufficiently rich and long so as to fully excite the system yielding an output sequence that is representative for the system's behavior.
\end{remark}

During the online implementation, at each time step $t$,  an input/output trajectory of the past $T_\mathrm{ini}$ steps is buffered and used to form $u_\mathrm{ini}=u_{[t-T_\mathrm{ini},t-1]}$ and $y_\mathrm{ini}=y_{[t-T_\mathrm{ini},t-1]}$. 
Let $u=u_{[t,t+N-1]}$ and  $y=y_{[t,t+N-1]}$ be the control inputs and the system outputs over the $N$-step prediction horizon, respectively. The Willems' fundamental lemma states that if the pre-collected input sequence $u^d_{[1,T]}$ is persistently exciting of order $T_\mathrm{ini}+N+n$ with $n$ being the dimension of the system states ($n$ can be chosen as an upper bound of state dimension \cite{DeepC}), then the patched trajectory $(u_\mathrm{ini},y_\mathrm{ini},u,y)$ is generated from the LTI system when it is spanned by $(U_{\mathrm{p}},Y_{\mathrm{p}},U_{\mathrm{f}},Y_{\mathrm{f}})$, that is, there exists a vector $g\in\mathbb{R}^{T-T_\mathrm{ini}-N+1}$ such that:
\begin{equation}\label{Eq:FundLemma}
    \begin{bmatrix}
	U_\mathrm{p} \\ Y_\mathrm{p} \\ U_\mathrm{f} \\ Y_\mathrm{f}
    \end{bmatrix}g=
    \begin{bmatrix}
	u_\mathrm{ini} \\ y_\mathrm{ini} \\ u \\ y
    \end{bmatrix},
\end{equation}
which is a non-parametric system representation that ensures any consecutive trajectory of length $T_\mathrm{ini}+N$ is compatible with those in the trajectory library in (\ref{eq:Hankel}), and the initial trajectory $[u_\mathrm{ini}^\top,y_\mathrm{ini}^\top]^\top$ can be viewed as setting an initial condition for the unknown future trajectory $[u^\top,y^\top]^\top$. The DeePC can then be cast as the following constrained optimization problem:
\begin{equation} \label{Eq:DeePC}
\begin{aligned}
\min_{g,u,y} & \quad J(u,y) = \sum\limits_{k=t}^{t+N-1}\left( \left\|y(k)\right\|_{Q}^{2}+\left\|u(k)\right\|_{R}^{2}\right)\\
\mathrm{s.t.} &\quad  \eqref{Eq:FundLemma},\\
& \quad y(k) \in\mathcal{Y}(k), \quad k=t, \ldots, t+N-1,\\
& \quad u(k) \in\mathcal{U}(k), \quad k=t, \ldots, t+N-1,
\end{aligned}
\end{equation}
where $\mathcal{U}(k)$ and $\mathcal{Y}(k)$   represent the input and output  constraints at time step $k$, respectively. As the DeePC bypasses the model identification step and directly optimizes the control from data, it has found great successes across various domains \cite{DeepC,DeePC-Motor,DeePC-PowerGrid,DeePC-CAV}.

\subsection{Regularized DeePC}
DeePC was first developed for deterministic LTI systems. For nonlinear systems, a regularization scheme has been developed to achieve satisfactory performance \cite{DeepC,DeePC-Stochastic,DeePC-Page,DeePC-Quadcopters}.
In particular, for the input/output data collected from a nonlinear system, the subspace spanned by the constructed Hankel matrix may no longer be consistent with the subspace of trajectories from the underlying system, even if the Hankel matrix satisfies the full rank condition under such data sets collected from a nonlinear system. A direct consequence is that poor prediction performance is likely to  happen with the ill-posed condition of Hankel matrix constraint. In this case, a penalty term is added for the discrepancy, quantified by a slack variable $\sigma_y$, between the estimated initial condition $Y_{\mathrm{p}}g$ and the real-time buffered initial condition $y_\mathrm{ini}$, which provides a least-square estimation of the true initial condition. Then with an additional regularization term, the corresponding regularized DeePC optimization problem is presented as:
\begin{subequations} \label{Eq:reg-DeePC}
    \begin{align}
        \min_{g,u,y} & \quad J(u,y) + \lambda_y||\sigma_y||_2^2 + \lambda_g||g||_2^2\label{Eq:reg_DeePC_cost}\\
        \mathrm{s.t.} &\quad
        \begin{bmatrix}
		    U_\mathrm{p}
		    \\Y_\mathrm{p}
		    \\ U_\mathrm{f}
		    \\ Y_\mathrm{f}
	    \end{bmatrix}g=
	    \begin{bmatrix}
		    u_\mathrm{ini}
		    \\ y_\mathrm{ini}
		    \\ u
		    \\ y
		\end{bmatrix}+
		\begin{bmatrix}
		    0
		    \\ \sigma_y
		    \\ 0
		    \\ 0
		\end{bmatrix},\label{Eq:reg-fundL}\\
        & \quad y(k) \in\mathcal{Y}(k), \quad k=t, \ldots, t+N-1,\\
        & \quad u(k) \in\mathcal{U}(k), \quad k=t, \ldots, t+N-1,
    \end{align}
\end{subequations}
where $\lambda_y, \lambda_g\in \mathbb{R}_+$ are the regularization parameters. This problem formulation \eqref{Eq:reg-DeePC} has been proved to coincide with a distributionally robust problem via the technique of Wasserstein metric \cite{DeePC-Stochastic,DeePC-Page}.

\begin{remark}[selection of $\lambda_y$]\label{remk:lmday}
    $\lambda_y$ can be  chosen as a scalar to equally penalize the  elements in $\sigma_y=Y_\mathrm{p}g-y_{\mathrm{ini}}$. Alternatively, $\lambda_y$ can be designed as a penalty matrix to transform the second term in \eqref{Eq:reg_DeePC_cost} to be $||\sigma_y||^2_{\lambda_y}$ so that  elements in $\sigma_y$ can be penalized differently by assigning the weighting factor in the corresponding position of matrix $\lambda_y$.
\end{remark}

\section{DeePC for Li-ion Battery Fast Charging}\label{sec:DeePC}
In this section, we first briefly introduce  the Li-ion battery system to provide background information on our considered application. Then, a non-parametric system representation is developed to characterize the charging dynamics directly based on input-output measurements, without the need for complicated battery models such as ECMs and EMs. Finally, we introduce the control objective and system constraints and present the full DeePC formulation for Li-ion battery charging.

\subsection{Description of Li-ion Battery System} \label{subsec:Battery}
The working mechanism of Li-ion batteries is vastly complicated, with coupled chemical, electrical and thermal dynamics closely interacting with each other to influence the battery states. Physical or chemical reactions with different materials \cite{Li-ion-Materials} take place at different regions inside the battery, leading to varying concentrations of lithium ions, including battery internal electrolyte concentration ($c_e$) and concentration inside or at the surface of solid material particles ($c_s^{in}/c_s^{suf}$), whose dependence on other battery states can be represented by the abstract partial differential equations:
\begin{equation}\label{eq:concentration}
    \begin{aligned}
        \frac{\partial c_{e,i}(x,t)}{\partial t}& + \mathcal{L}_1 c_{e,i}(x,t) = F_1(I(t),c_{s,i}(x,t),T_{emp}(x,t),\eta_i),\\
        \frac{\partial c_{s,i}(x,t)}{\partial t}& + \mathcal{L}_2 c_{s,i}(x,t)= F_2(I(t),c_{e,i}(x,t),T_{emp}(x,t),\eta_i),
    \end{aligned}
\end{equation}
where $\mathcal{L}_1$ and $\mathcal{L}_2$ are operators acting on the abstract space of $c_{e}$ and $c_s$ for their spatial distributions. $F_1(\cdot)$ and $F_2(\cdot)$ are abstract nonlinear reaction functions that describe the intrinsic spatial-temporal relation of the lithium-ions inside the battery (see \cite{electrochemical-doyle, battery_system} for the detailed forms). Here $i\in\mathcal{S}$ represents the index of the battery region, where $\mathcal{S}=\{anode,separator, cathode\}$. For detailed discussions about the battery region separation, interested readers can refer to \cite{Alg4BMS}.  $I$ and  $T_{emp}$ represent the charging 
current and the battery temperature, respectively. Furthermore, $\eta$ characterizes the over-potential of interface reactions and can be represented by \cite{electrochemical-ZoneMPC,LionSimba}:
\begin{equation}
    \eta_i=\Phi_s(x,t)-\Phi_e(x,t)-U_i,
\end{equation}
where $\Phi_e$ denotes the electrolyte potential on one side of the interface whereas $\Phi_s$ denotes the solid material potential on the other side; and $U_i$ denotes the open circuit (or equilibrium) potential. The dependence of $\Phi_e$ and $\Phi_s$ on battery internal states can similarly be described in form of \eqref{eq:concentration} as,
\begin{equation}\label{eq:potential}
    \begin{aligned}
        \frac{\partial \Phi_s(x,t)}{\partial t}& + \mathcal{C}_1\Phi_s(x,t)= G_1\big(I(t),c_{e,i}(x,t),c_{s,i}(x,t),T_{emp}(x,t)\big),\\
        \frac{\partial \Phi_e(x,t)}{\partial t}& + \mathcal{C}_1\Phi_e(x,t)= G_2\big(I(t),c_{e,i}(x,t),c_{s,i}(x,t),T_{emp}(x,t)\big),
    \end{aligned}
\end{equation}
where $\mathcal{C}_1$ and $\mathcal{C}_2$ are also the abstract operators, operating on $\Phi_s$ and $\Phi_e$, respectively. $G_1(\cdot)$ and $G_2(\cdot)$ abstractly describe the battery internal relations among all states.
\begin{comment}
where $G_1(\cdot)$ and $G_2(\cdot)$ are also unknown, while the example function references refer to \cite{electrochemical-doyle, battery_system}.
\end{comment}
The aforementioned function formats can generally take the form of partial differential-algebraic equations \cite{electrochemical-doyle,electrochemical-P2D,electrochemical-SPM,electrochemical-Segment,electrochemical-NMPC}, which are  complex and tightly coupled. Furthermore,   comprehensive  Li-ion battery models also take into account  the temperature variation driven by irreversible and reversible heat generations, with the following heat equation:
\begin{equation}\label{eq:thermalDyn}
    \frac{\partial T_{emp}(x,t)}{\partial t} + \mathcal{D}T_{emp}(x,t)= Q_{irr}\big(I(t)\big)+Q_{rev}\big(I(t)\big).
\end{equation}
Here the operator $\mathcal{D}$ characterizes the spatial gradient of the temperature in the battery; $Q_{irr}$ represents the irreversible heat generation \cite{heatsource_irr1,heatsource_irr2,heatsource_irr3,heatsource_irr4}, with resistive heating as a major source of contribution; and $Q_{rev}$ represents the reversible heat process, also known as entropic heat \cite{heatsource_rev}.

In addition, boundary conditions in the above PDEs  are required on the aforementioned battery states (e.g., $c_e,c_s,\Phi_e,\Phi_s,T_{emp}$) for a physically meaningful solution, i.e., all electrical and chemical reactions are limited within certain spatial regions of the battery and heat exchanges within the surrounding environment \cite{Alg4BMS}. Furthermore, interface conditions \cite{Alg4BMS,LionSimba} are also required to enforce the continuity of the solution between two different materials inside the battery.
\begin{comment}
so that more intermediate equations \cite{Alg4BMS,LionSimba} are needed to be evaluated at the interfaces between two different materials inside the battery.
\end{comment}
It is obvious that the battery charging dynamics is very complicated and for control purposes, ECM and EM methods are typically  reduced to a set of ordinary differential equations (ODEs), 
\begin{equation}\label{eq:battery_ode}
    \begin{aligned}
        \dot{z}(t) &=f(z(t),u(t)),\\
        y(t)&=h(z(t),u(t)),
    \end{aligned}
\end{equation}
where $f(\cdot)$ denotes the simplified model from \eqref{eq:concentration} and \eqref{eq:thermalDyn}, and $h(\cdot)$ is the output function. The system state $z$ is chosen as a subset of the battery states (e.g., $c_e,c_s,\Phi_e,\Phi_s,T_{emp}$), and $u$ is generally chosen as the charging current $I$. The control objective is to obtain an admissible control inputs $u(t)\in\mathcal{U}(t)$ such that when applied to system \eqref{eq:battery_ode}, the resulting output $y(t)\in\mathcal{Y}(t)$ is also admissible while minimizing a predefined cost function to achieve safe and fast battery charging. More formally, the optimization problem can be formulated as,
\begin{equation}\label{Eq:general_Obj}
    \begin{aligned}
	\min_{u,y} \quad &\int_{t}c(u(\tau),y(\tau))d\tau\\
        \text{s.t.} \quad &\quad \eqref{eq:battery_ode}\\
        & u(t)\in\mathcal{U}(t),\quad y(t)\in\mathcal{Y}(t).\\
    \end{aligned}
\end{equation}
Here $c(\cdot)$ denotes the running cost, starting from time instant $t$ and integrated over a time period for the specified battery charging cycle. For example, several MPC frameworks have been proposed \cite{electrochemical-NMPC,electrochemical-ZoneMPC,electrochemical-QDMC,equC-MPC}. However, it is clear from the above discussions that such model-based approaches are very challenging to design due to the exceedingly difficult modeling and calibration. Therefore, we next present an efficient data-enabled approach for effective controls by directly exploiting input-output data, without the need for the complex modeling process.

\begin{figure*}[h]
  \centering
  \includegraphics[width=0.5\textwidth]{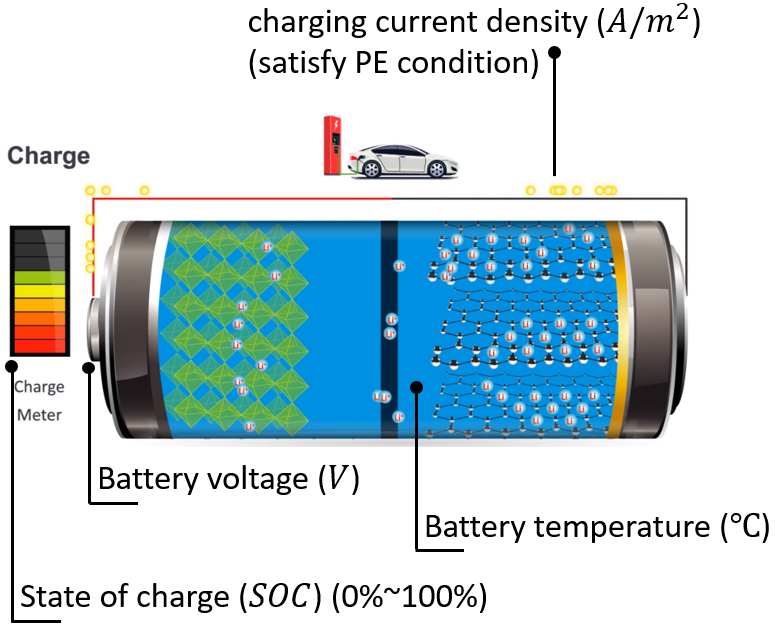}
  \caption{{\small Measurable inputs and outputs of Li-ion batteries.}}\label{fig:Battery}
\end{figure*}

\subsection{Non-Parametric Representation of Li-Ion Battery System for Fast charging}
In this section, we present a non-parametric, data-driven system representation that directly describes the battery dynamics using input-output data. Specifically, we exploit the behavioral system theory and Willems' fundamental lemma to characterize the system using the input-output data as detailed in Section~II-B. More specifically, as shown in Figure~\ref{fig:Battery}, the system outputs are specified as battery voltage ($y_V$), battery temperature ($y_{Temp}$), and state of charge ($y_{soc}$) while the system input is chosen as the charging current ($I$). Then an input-output trajectory of length $T$ is collected by exciting the system with an input trajectory that is PE of order $L$. The inputs can be designed as sinusoidal signals or multi-level pseudo-random signals \cite{PRML}. As a result, we obtain the following data vectors:
\begin{equation*}
    \begin{aligned}
        u^d_{\left[1, T\right]}=\begin{bmatrix}
        u^d(1) \\ \vdots \\ u^d(T)
        \end{bmatrix} \in \mathbb{R}^{mT}, \quad y^d_{\left[1, T\right]}=\begin{bmatrix}
        y^d(1) \\ \vdots \\ y^d(T)
        \end{bmatrix} \in \mathbb{R}^{rT},
    \end{aligned}
\end{equation*}
with $u^d_{\left[1, T\right]}$ being the input charging current ($I$) sequence and $y^d_{\left[1, T\right]}$ the output sequence whose element $y(i)=[y_{V}(i),y_{T_{emp}}(i),y_{soc}(i)]^\top, i\in[1,T]$ consists of battery voltage ($V$), battery temperature ($T_{emp}$) and SOC. Note that even though SOC cannot be directly measured, reliable methods exist for accurate estimations \cite{Alg4BMS}. Therefore, the choice of input-output data is practically viable. The pre-collected input-output data vector is then re-arranged into the Hankel matrices \eqref{eq:Hankel} for non-parametric representation of battery charging dynamics, with partitioned blocks corresponding to the ``past'' data of length $T_\mathrm{ini}$ and ``future'' data of length $N$.

Finally, during online implementation, $u_\mathrm{ini}=u_{[t-T_\mathrm{ini},t-1]}$ and $y_\mathrm{ini}=y_{[t-T_\mathrm{ini},t-1]}$ of length $T_\mathrm{ini}$  are buffered. Define $u=u_{[t,t+N-1]}$ and  $y=y_{[t,t+N-1]}$ of length $N$ as the control sequence and output sequence in the optimization horizon of length $N$, the system behavior can be represented by the non-parametric model \eqref{Eq:reg-fundL}, which circumvents the major challenges in deriving and calibrating the complicated EMs and ECMs.

\begin{remark}[Construction of Hankel matrix]\label{remk:HankelM}
    If it is challenging for the system to generate a long trajectory, e.g., the system has unstable dynamics or it is expensive to run a long test. In this case, multiple short system trajectories can be patched to construct the Hankel matrix, as long as it satisfies a collective persistency of excitation condition \cite{Fundamental-Lemma-MultiDataSet}.
\end{remark}

\subsection{DeePC Formulation for Li-ion Battery Fast Charging }
The control goal here is to charge the Li-ion battery as fast as possible, while ensuring system safety. Specifically, the objective is to design an optimal charging current profile to complete the charging task with minimum time consumed, while satisfying safe working condition constraints. To facilitate the practical application, instead of using the total time as a cost during the whole charging process, the objective function is designed by driving SOC to a target value. In addition, to alleviate the current chattering phenomenon and maintain a smooth charging process, the input variation in successive optimization steps are penalized \cite{electrochemical-ZoneMPC, electrochemical-QDMC,equC-MPC2}. As a result, the cost function is designed as:
\begin{equation} \label{Eq:BFC_cost}
J(u,y) = \sum\limits_{k=t}^{t+N-1}\left( \left\|y_{soc}(k)-r_{soc}\right\|_{Q}^{2}+\left\|\Delta u(k)\right\|_{R}^{2}\right),\\
\end{equation}
where $Q$ and $R$ are penalty matrices. By adding the regularization terms for DeePC extension to account for system nonlinearity, the cost function is reduced to the form of \eqref{Eq:reg_DeePC_cost}.

To ensure safety during charging, in this paper,  extreme charging scenarios are restricted: First, the charging rate will be maintained under $2C$\footnote{C-rate: it indicates the ratio of current magnitude to the battery size (battery nominal capacity).}, because rapid heat generation \cite{electrochemical-Segment} is generally accompanied by high charging rate, causing battery degradation, and terminal voltage will also increase fast \cite{electrochemical-ZoneMPC} to cut-off value, leading to an early termination of battery charging; Second, the limits of lower and upper cut-off voltages are considered, and if the limit is reached then the charging is stopped to avoid over-powered charging beyond safe charging zone \cite{equC}; and Third, the battery temperature is constrained between $25^\circ C$ and $31^\circ C$ to avoid accelerated battery degradation at high temperatures \cite{electrochemical-Segment,electrochemical-NMPC,BatteryAging}. All these considerations are geared towards demonstrating the benefits of DeePC for Li-ion battery fast and safe charging purposes, without considering negligible battery internal side reactions. In the long run, however, the side reactions may cause irresistible degradation to the Li-ion battery, and lots of works are focusing on the avoidance and/or alleviation of such side reactions. Safe and fast charging of Li-ion battery with consideration of side reaction effects is still an open research field and will be considered in our future work.

To sum up, to maintain safe operations during charging, the following constraints are enforced:
\begin{equation} \label{Eq:BFC_constraints}
	\begin{aligned}
	    I_{min} \leq &u(k) \leq I_{max},\\
		\Delta I_{min} \leq &\Delta u(k) \leq \Delta I_{max},\\
		V_{min}  \leq & y_{V}(k) \leq V_{max},\\
		T_{emp,min} \leq &y_{T_{emp}}(k) \leq T_{emp,max},\\
		SOC_{init} \leq &y_{soc}(k) \leq SOC_{final},\\
	\end{aligned}
\end{equation}
where $I_{min/max}$ and $\Delta I_{min/max}$ denote the lower and upper bounds of input charging current and its fluctuation, respectively; $V_{min/max}$ denotes the cut-off voltage for low/high battery voltage bounds; $[T_{emp,min},T_{emp,max}]$ denotes the desired battery working temperature range; and $SOC_{init}$ and $SOC_{final}$ are the desired starting point and  ending point for battery charging, respectively. The specific values will be shown in Section \ref{sec:simulation}.

\section{DeePC with Dimension Reduction}\label{sec:eff-DeePC}
Despite promising performance of (regularized) DeePC demonstrated in various applications, its computational complexity is high as it solves the optimization problem (\ref{Eq:reg-DeePC}) with a large dimension of optimization variables $g\in\mathbb{R}^{T-T_{ini}-N+1}$, where $T$ is often much larger than ${T_{ini}+N}$ to satisfy the persistent excitation (PE) condition. As a result, DeePC is generally more computationally expensive than its MPC counterpart. In this section, we  exploit PCA \cite{SVD} based method for dimension reduction to facilitate fast optimization while retaining the performance. 

Specifically, we denote the left-side data matrix of \eqref{Eq:FundLemma} as $A$ and the right-side date vector of \eqref{Eq:FundLemma} as $b$:
\begin{equation}\label{Eq:Ab}
	{A}:=\begin{bmatrix}
			U_\mathrm{p} \\Y_\mathrm{p} \\ U_\mathrm{f} \\ Y_\mathrm{f}
		\end{bmatrix},\qquad
	b :=\begin{bmatrix}
			u_\mathrm{ini} \\ y_\mathrm{ini} \\ u \\ y
		\end{bmatrix}.
\end{equation}
Note that in order to satisfy the PE condition and attain good performance, a large number of columns in ${A}$ are generally used \cite{DeepC,DeePC-CAV}. Since the fundamental principle is that the columns of ${A}$ must span the $T_{ini}+N$ continuous input/output trajectories, we  employ PCA to identify the $l$ most important representative trajectories, which we refer to as \emph{eigen-trajectories}. Here $l$ is a hyper-parameter that can be tuned. More specifically, the following dimension reduction steps are performed:

\begin{enumerate}[1)]
\item Let $p=(m+r)(T_{ini}+N)$ and $q=T-T_{ini}-N+1$, then the singular value decomposition (SVD) of matrix $A$ is performed, as follows: 
\begin{equation}
    {A} = W\Sigma V^\top,
\end{equation}
where matrix $\Sigma\in \mathbb{R}^{p\times q}$ is a rectangular diagonal matrix and uniquely determined by $A$ with all the singular values $\sigma_i=\Sigma_{ii}, i\leq \text{min}\{p,q\}$ on the diagonal in descending order. The columns of matrix $W\in \mathbb{R}^{p\times p}$ and the columns of matrix $V \in \mathbb{R}^{q\times q}$ are called left-singular vectors and right-singular vectors of $A$, respectively. They also form two sets of orthogonal bases.

\item Choose the first $l$ columns from $V$ to form a new matrix which is denoted as $V_{[1:l]}$, where $l\leq q$. As the columns of $V$ represent the principle mixture of system modes \cite[C. 1]{DD_book} evolving in time, the first $l$ columns of $V$ are the corresponding most representative trajectories, which we call \emph{eigen-trajectories}, and they are essentially utilized for mapping.

\item Define a new vector $\bar{g}$ with the relationship to $g$ as,
\begin{equation}
    g=V_{[1:l]}\bar{g},
\end{equation}
where $\bar{g}$ has the dimension $l$. The matrix $V_{[1:l]}$ here, as mentioned, is a mapping from $\bar{g}$ to $g$.

\item The equation \eqref{Eq:FundLemma} in the DeePC problem can be replaced by
\begin{equation}
    \bar{A}\bar{g}=b,
\end{equation}
where $\bar{A}=W\Sigma V^\top V_{[1,l]}=AV_{[1,l]}$, and the optimization variable changes from $g$ to $\bar{g}$ with a lower dimension. The same equation transformation and replacement can be applied to \eqref{Eq:reg-fundL}.
\end{enumerate}

\begin{remark}[Selection of Column Length $l$]\label{remk:Lselection}
    Three methods can be used to choose $l$ for truncation of right-singular matrix $V$ to get $V_{[1,l]}$,
    \begin{itemize}
        \item Principal component spectrum analysis. Plot the main modes in descending order, $l$ can be picked by identifying the 'elbow' point with the steepest slope or some other points with the required energy percentage along the curve.
        \item Cross-validation. it is a kind of trial-and-error method by evaluating a performance index defined based on the variation of the length $l$.
        \item Optimal Hard Threshold Method \cite{SVD-optimalTrunc}. An optimal mode truncation point can be calculated by finding the underlying lower rank matrix, especially in cases under heavy noise corruption.
    \end{itemize}
\end{remark}

\section{Simulations}\label{sec:simulation}
In this section, the DeePC approach for Li-ion battery fast charging problem with safety constraints is implemented via simulation to demonstrate its effectiveness. In addition, comparison is also made for the proposed PCA-based dimension reduction strategy on DeePC, showing its advantages of promoting online optimization efficiency and, meanwhile, still retaining the optimal performance of DeePC without dimension reduction.

\subsection{Experimental Setup} \label{subsec:setup}
A high-fidelity battery simulation model - LIONSIMBA \cite{LionSimba}, is used as the plant for simulation studies. The LIONSIMBA simulator is built based on the first principles and can provide all battery dynamics related signals, so the aforementioned signals, i.e., battery voltage, temperature, SOC, and charging current, are measurable or estimated as those from a real Li-ion battery management system.

The offline data $(u^d,y^d)$ is collected with the sampling interval chosen as $\Delta t=10s$. Among this collected data set, the input data sequence $u^d$ is designed as a combination of sinusoidal signals and multi-level pseudo-random signals for system excitation. The `past' data length and `future' data length are $T_\mathrm{ini}=60$ and $N=70$, respectively. In the cost function of DeePC problem, only one of the three outputs (SOC) is penalized, with the corresponding weighting factor in $Q\in\mathbb{R}^{3\times 3}$ selected as 10, and the weighting factor for charging current is chosen as $R=0.1$. The coefficients of regularization terms are set as $\lambda_g=10^5, \lambda_{y_{soc},y_V}=10^3$ and $\lambda_{y_T}=10^6$. The lower and upper bounds for system constraints are listed in Table~\ref{tab:param}. Simulations are conducted using Matlab R2021b on Windows 10@3.6GHz PC with 8GB RAM. Before every simulation run, an ambient temperature $T_{amb}=25^\circ C$ is set as the initial condition of the simulator, and the charging current is selected as $I=0C$. Note that for the chemistry property of this simulator, $1C$ value is approximately $30A/m^2$. 

\begin{table}[!t]
\caption{\small{Safety constraint values}}
\label{tab:param}
\centering
\begin{tabular}{c c c c c}
%\hline
\hline
 $I_{min}$ & $\Delta I_{min}$ & $V_{min}$ $[V]$ & $T_{min}$ $[^\circ C]$ & $SOC_{init}$ $[\%]$\\
\hline
0 & $-0.1667C$ & 2.1 & 25 & 5\\
\hline
\hline
 $I_{max}$ & $\Delta I_{max}$ & $V_{max} $ $[V]$ & $T_{max}$ $[^\circ C]$ & $SOC_{final}$ $[\%]$\\
 \hline
$2C$ & $0.1667C$ & 4.17 & 31 & 95\\
\hline
%\hline
\end{tabular}
\end{table}

\begin{figure*}[t]
\centering
  \begin{subfigure}[c]{0.35\textwidth}
    \centering\includegraphics[width=\textwidth]{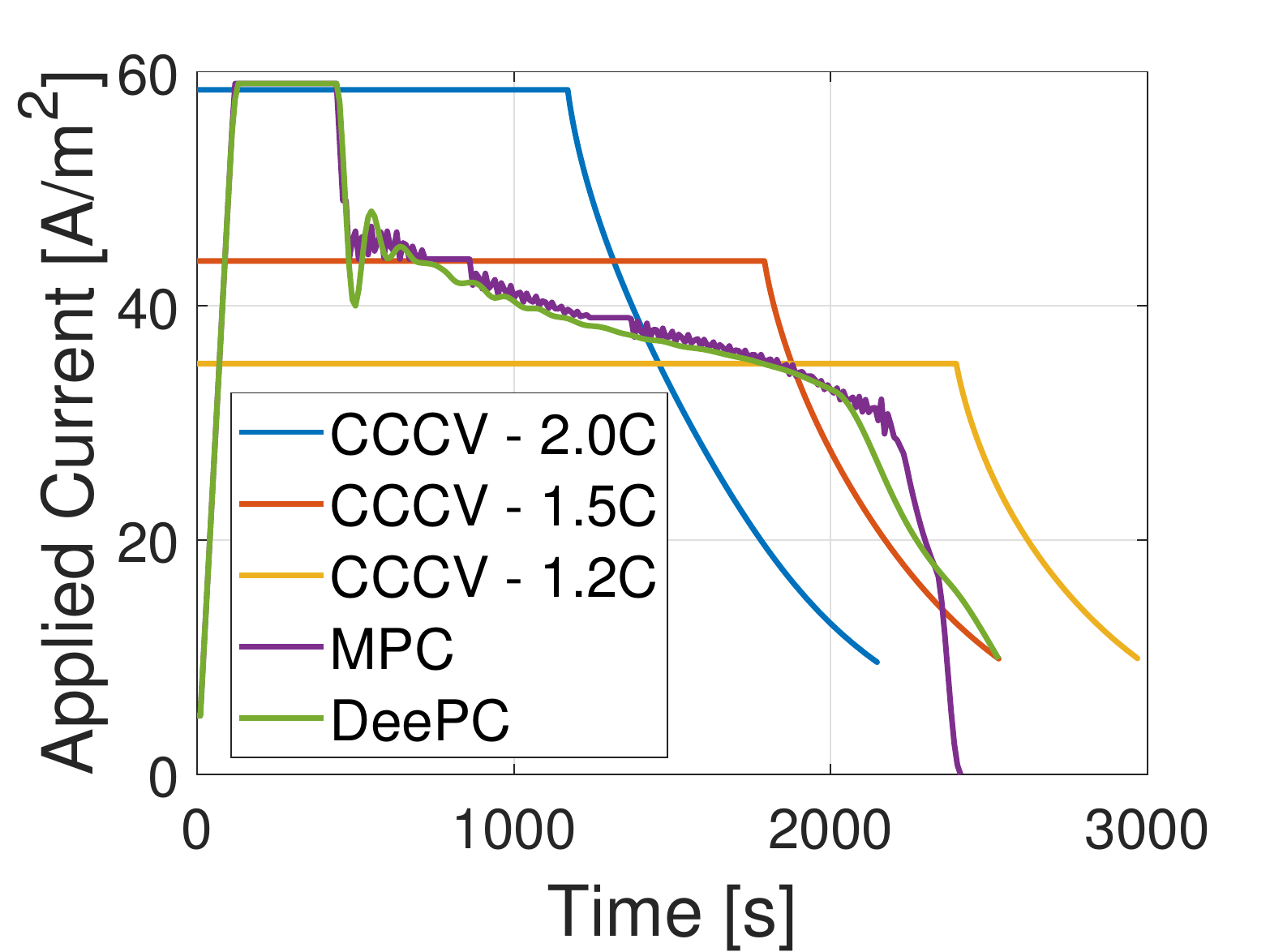}
    \caption{}
  \end{subfigure}
  \begin{subfigure}[c]{0.35\textwidth}
    \centering\includegraphics[width=\textwidth]{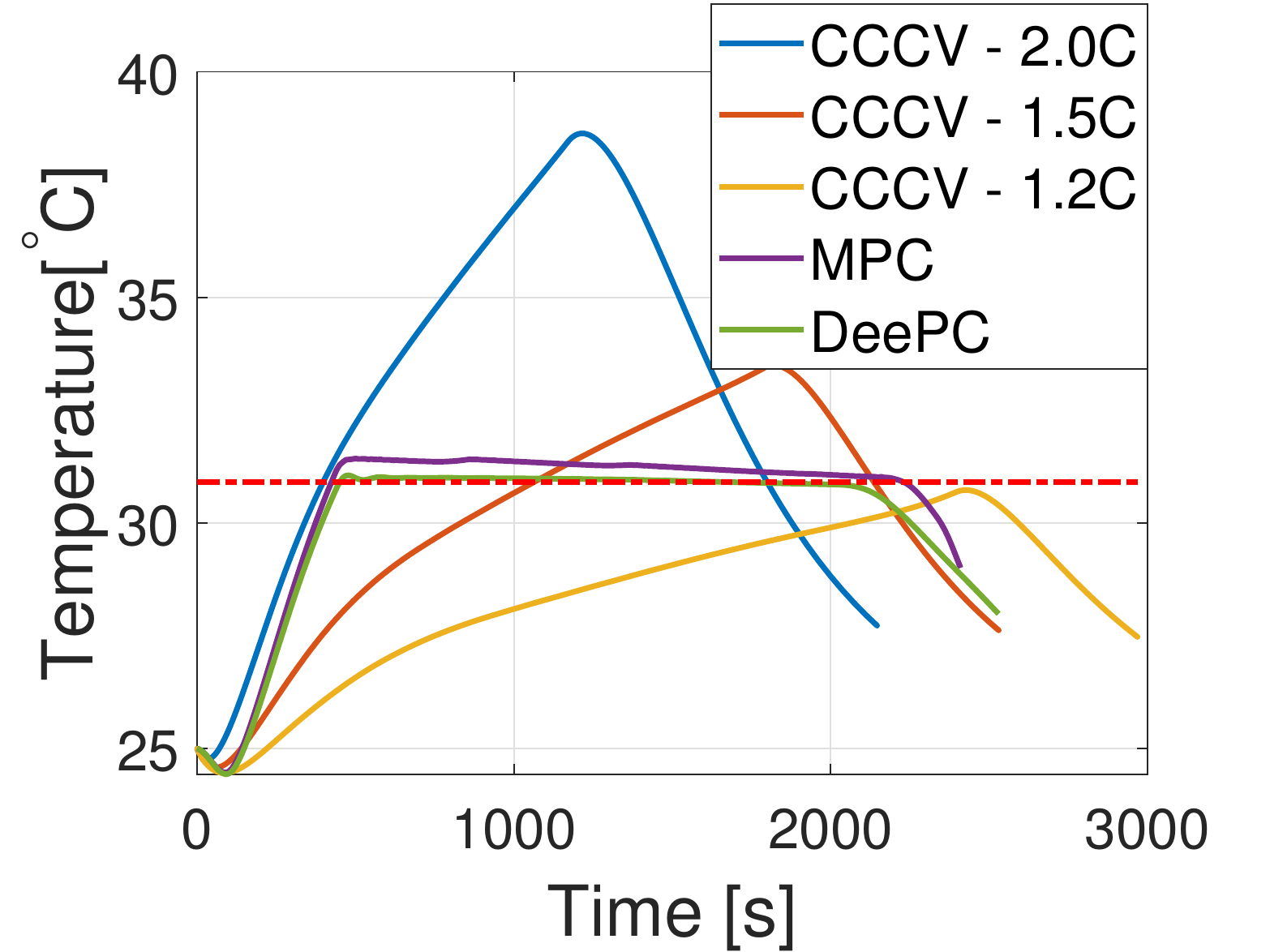}
    \caption{}
    \label{fig:DeePC_BFC_b}
  \end{subfigure}
 
  \begin{subfigure}[c]{0.35\textwidth}
    \centering\includegraphics[width=\textwidth]{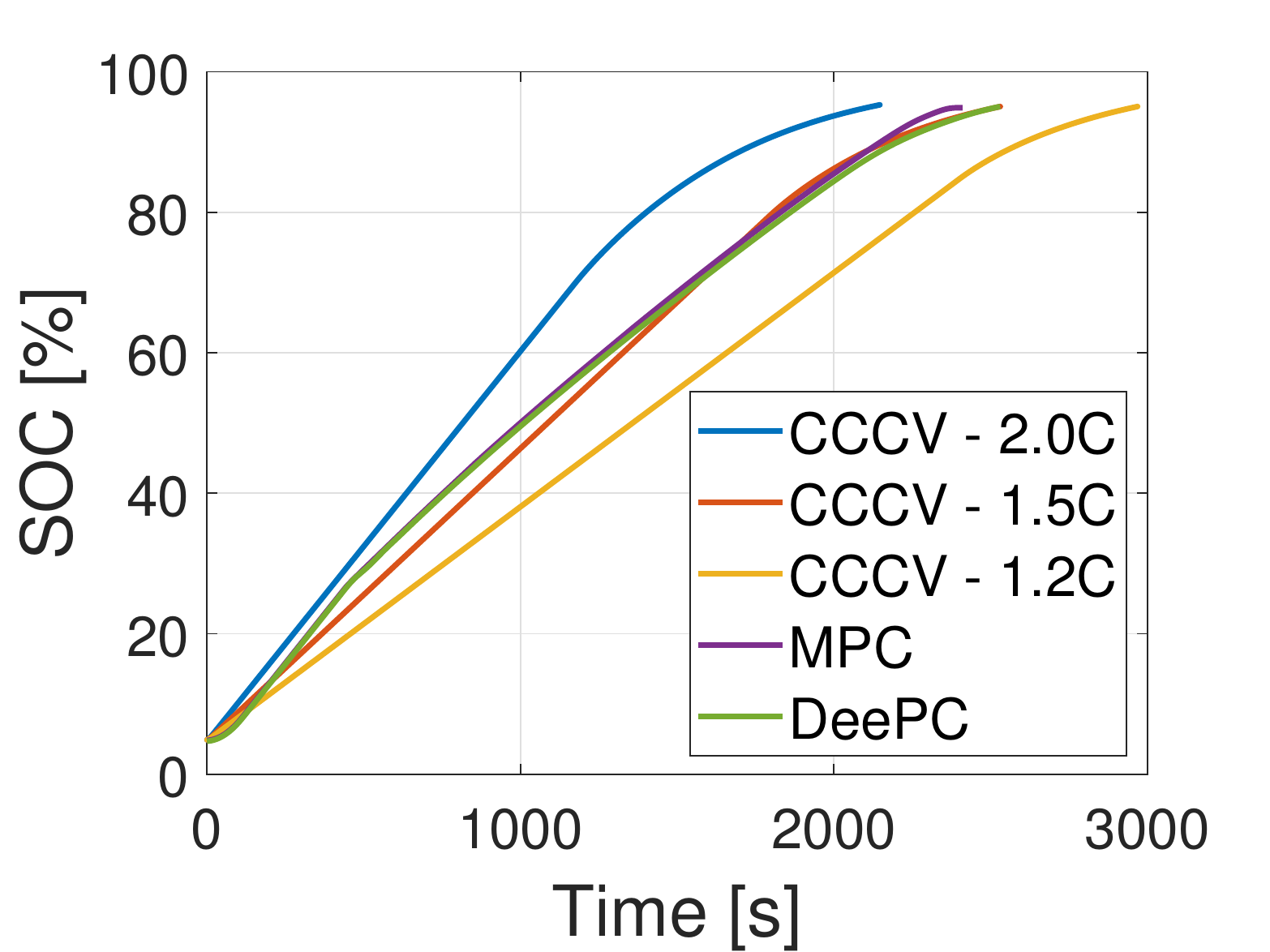}
    \caption{}
  \end{subfigure}
  \begin{subfigure}[c]{0.35\textwidth}
    \centering\includegraphics[width=\textwidth]{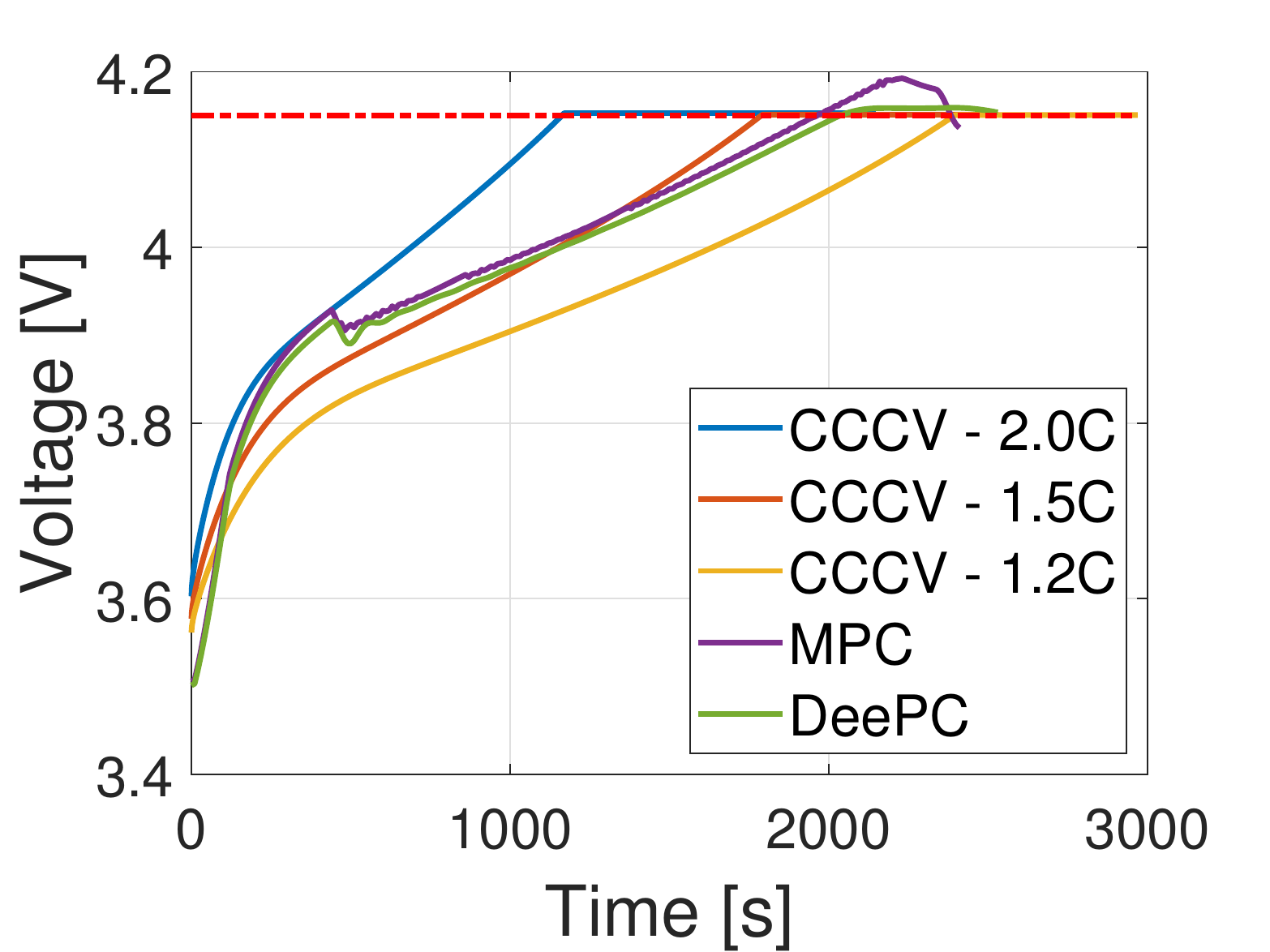}
    \caption{}
    \label{fig:DeePC_BFC_d}
  \end{subfigure}
  \caption{\small{Simulation results of DeePC and CCCVs.}}
  \label{fig:DeePC_BFC}
\end{figure*}

\subsection{Performance of DeePC for Li-ion Battery Fast Charging}
In this simulation, a fast charging target is set to charge the Li-ion battery from $5\%$ to $95\%$ SOC, almost a full cycle charge of the Li-ion battery. Both the CC-CV charging protocol and an MPC method are also implemented to facilitate the comparison. Specifically, three CC-CV protocols are applied for Li-ion battery charging, where the pre-set cut-off voltage value is $V_{cutoff}=4.15V$ and the constant charging currents are selected as $1.2C$, $1.5C$, and $2.0C$, respectively. In CC-CV protocols, the battery is initially charged in CC mode until the battery voltage reaches the cut-off value, and then the remaining charging proceeds in CV mode until SOC reaches $95\%$.
Moreover, for the MPC method, an auto-regressive with exogenous input (ARX) model is first estimated with the data set $(u^d,y^d)$ (see Section \ref{subsec:setup} for more details about the data collection) to capture the battery charging dynamics, and then the MPC problem is formulated with the ARX model and the safety constraints considered in the DeePC scheme.     

The simulation results are shown in Figure~\ref{fig:DeePC_BFC}. As CC-CV cannot handle system constraints, constraint violations occur for CC-CV-$1.5C$ and CC-CV-$2.0C$ protocols as depicted in Figure~\ref{fig:DeePC_BFC_b}. The temperatures rise above the safe zone during battery charging, and it is hazardous to apply such charging protocols. To potentially overcome this problem, the CC-CV schemes are often conservatively designed at the sacrifice of charging speed. 
%A typical example is the CC-CV-$1.2C$ charging protocol, which is deliberately designed to just satisfy all constraints, but the charging speed is the slowest, which takes $2970s$ for the SOC to rise from $5\%$ to $95\%$.
A typical example is the CC-CV-1.2C charging protocol, which is deliberately designed to satisfy all constraints. Nevertheless, the charging speed is the slowest, taking $2970$ seconds to charge the SOC from $5\%$ to $95\%$.  
Furthermore, it can be seen from Figrues~\ref{fig:DeePC_BFC_b} and \ref{fig:DeePC_BFC_d} that the MPC method oversteps the temperature and voltage constraints. As discussed in Section~\ref{subsec:Battery}, the battery charging dynamics are highly nonlinear and complicated, making it difficult to identify a model that can fully reflect the system features. Because the performance of MPC is heavily reliant on the estimated model, the modeling error will result in performance degradation such as constraint violations.

\begin{figure*}[t]
\centering
  \begin{subfigure}[c]{0.36\textwidth}
    \centering\includegraphics[width=1\textwidth]{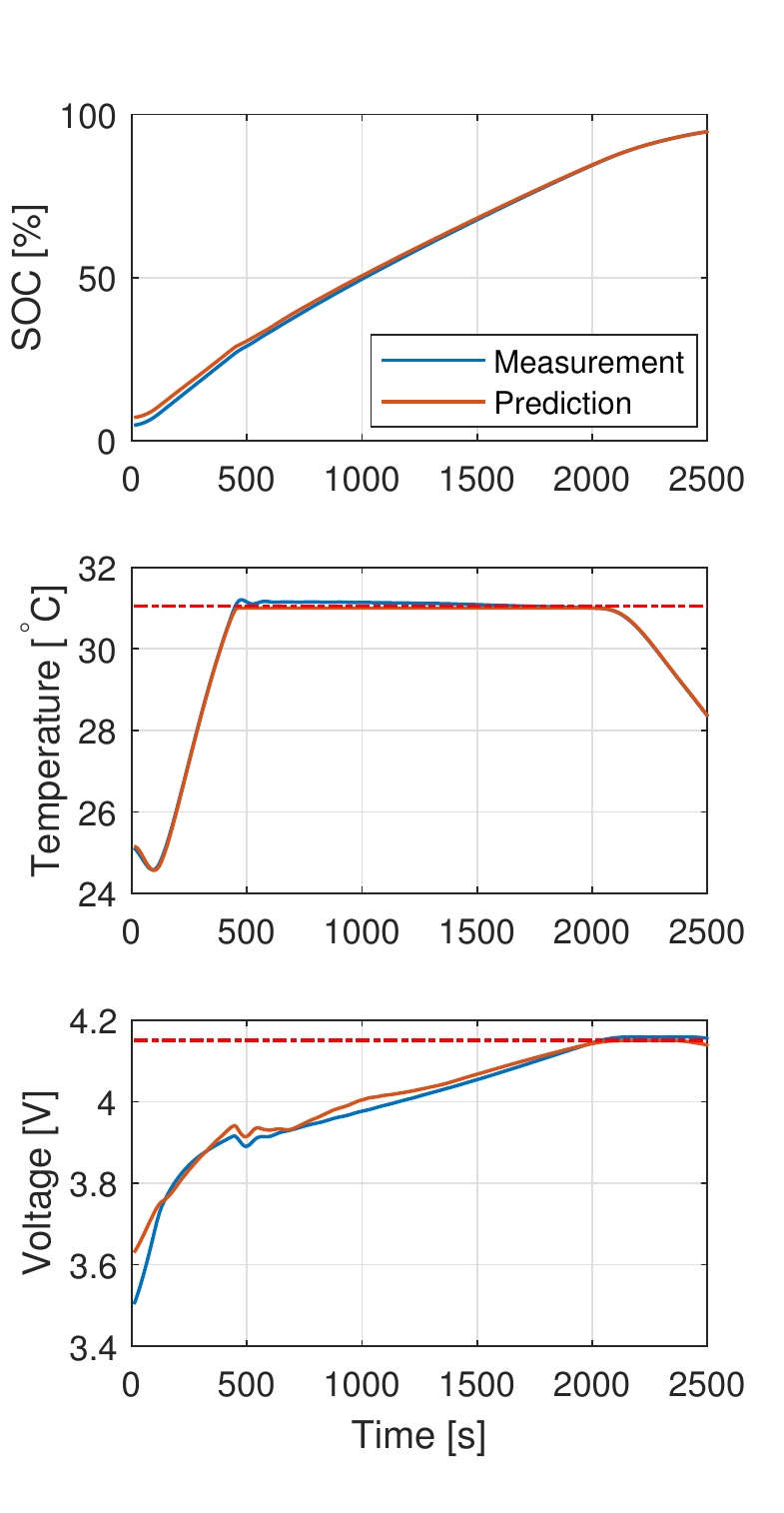}
    \caption{}
    \label{fig:pred_vs_meas_a}
  \end{subfigure}
  %\hfil
  \begin{subfigure}[c]{0.36\textwidth}
    \centering\includegraphics[width=1\textwidth]{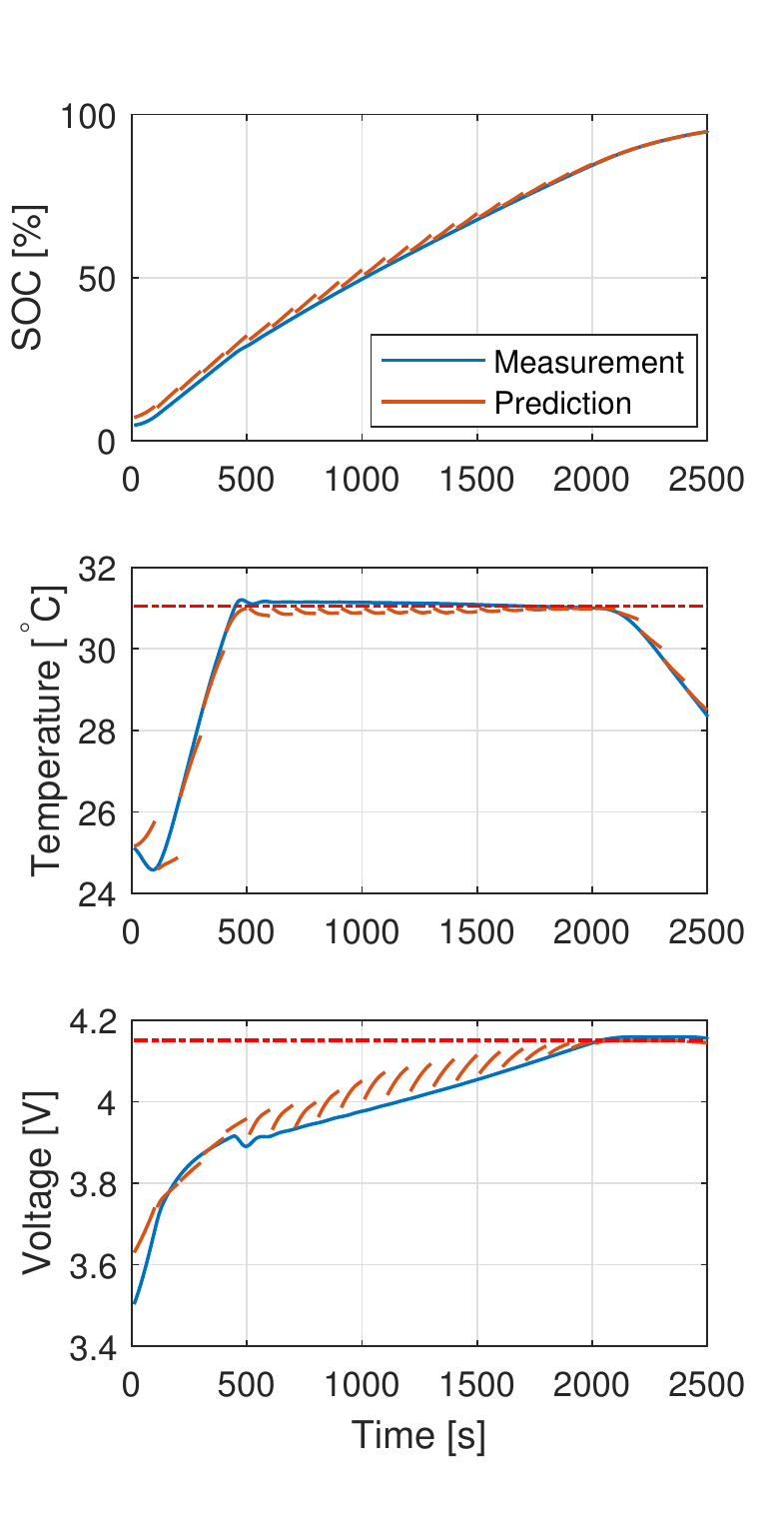}
    \caption{}
    \label{fig:pred_vs_meas_b}
  \end{subfigure}
  \caption{\small{DeePC prediction and real system measurement. (a) Comparison between $y_{1|t}$ and $y(t+1),t=0,1,2,\cdots$. (b) Comparison between $[y_{1|\bar{t}},\cdots,y_{10|\bar{t}}],\bar{t}=0,10,20,\cdots$ and $y(t+1),t=0,1,2,\cdots$.}}
  \label{fig:pred_vs_meas}
\end{figure*}

In contrast, the DeePC is a data-driven optimal controller with the ability to deal with system constraints. As also shown in Figure~\ref{fig:DeePC_BFC}, the DeePC not only satisfies all safety constraints, but also has a  smoother charging current profile. 
%The charging speed ($2530s$) of DeePC is almost the same as CC-CV-$1.5C$ ($2531s$ but it violates the constraints) and has an approximate $15\%$ improvement when compared with CC-CV-$1.2C$. 
DeePC's charging speed (2530 seconds) is nearly the same as CC-CV-$1.5C$ (2531 seconds but it violates the constraints) and is approximately $15\%$ faster than CC-CV-$1.2C$.
In addition, to illustrate the prediction accuracy of DeePC for battery fast charging, its prediction performances are compared with real system measurements at the same time instant. As shown in Figure~\ref{fig:pred_vs_meas_a}, the 1-step prediction performances of the three outputs match well with the system measurements. For the 10-step prediction case, shown in Figure~\ref{fig:pred_vs_meas_b}, even if certain deviations between successive prediction trajectories and the real measured ones exist, the overall trends of DeePC output prediction still have a good match to actual outputs. As a result, DeePC shows promising performance for Li-ion battery fast and safe charging while being simple and easy to implement.

\begin{figure*}[t]
\centering
  \begin{subfigure}[c]{0.35\textwidth}
    \centering\includegraphics[width=\textwidth]{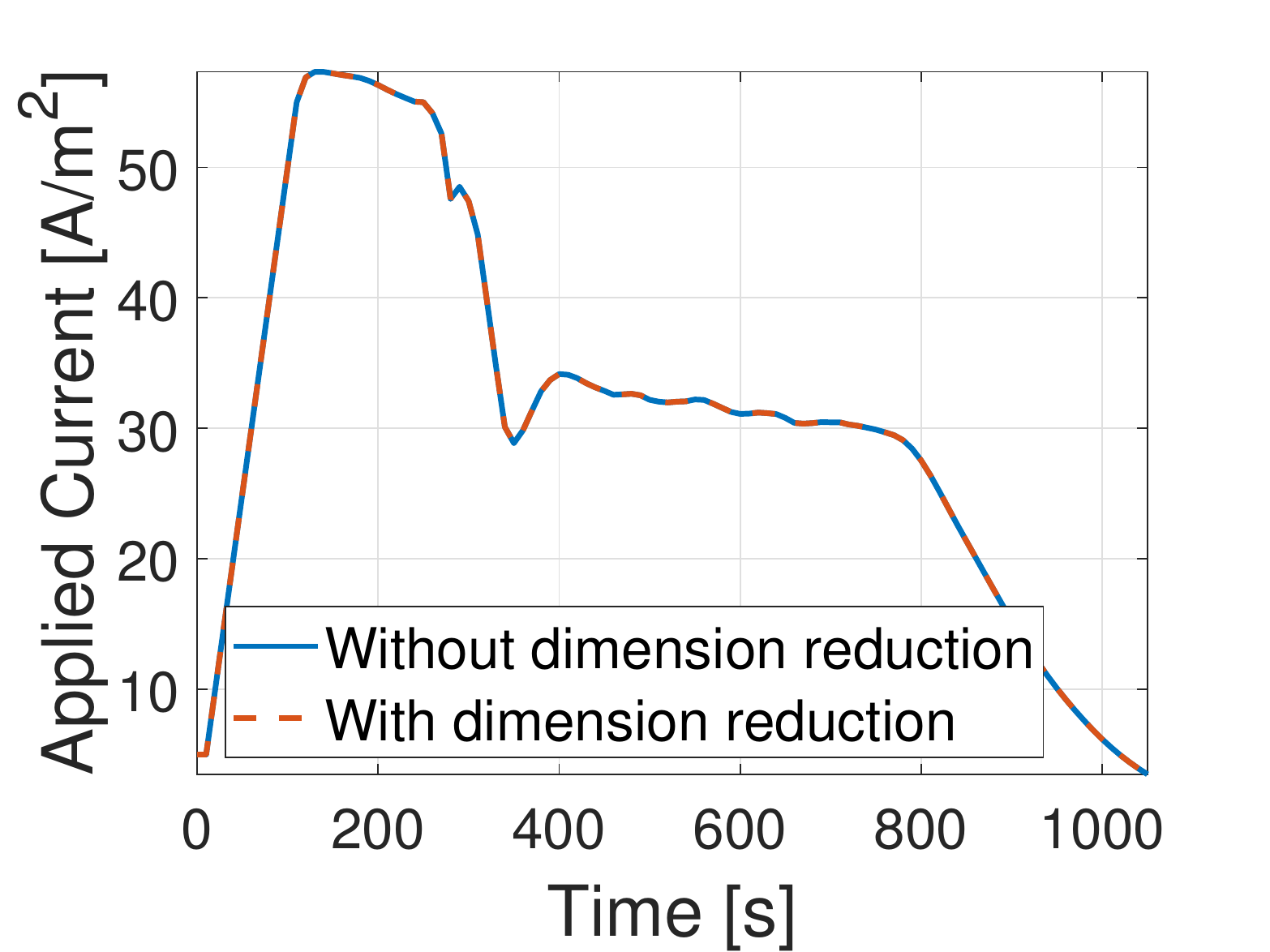}
    \caption{}
  \end{subfigure}
  \begin{subfigure}[c]{0.35\textwidth}
    \centering\includegraphics[width=\textwidth]{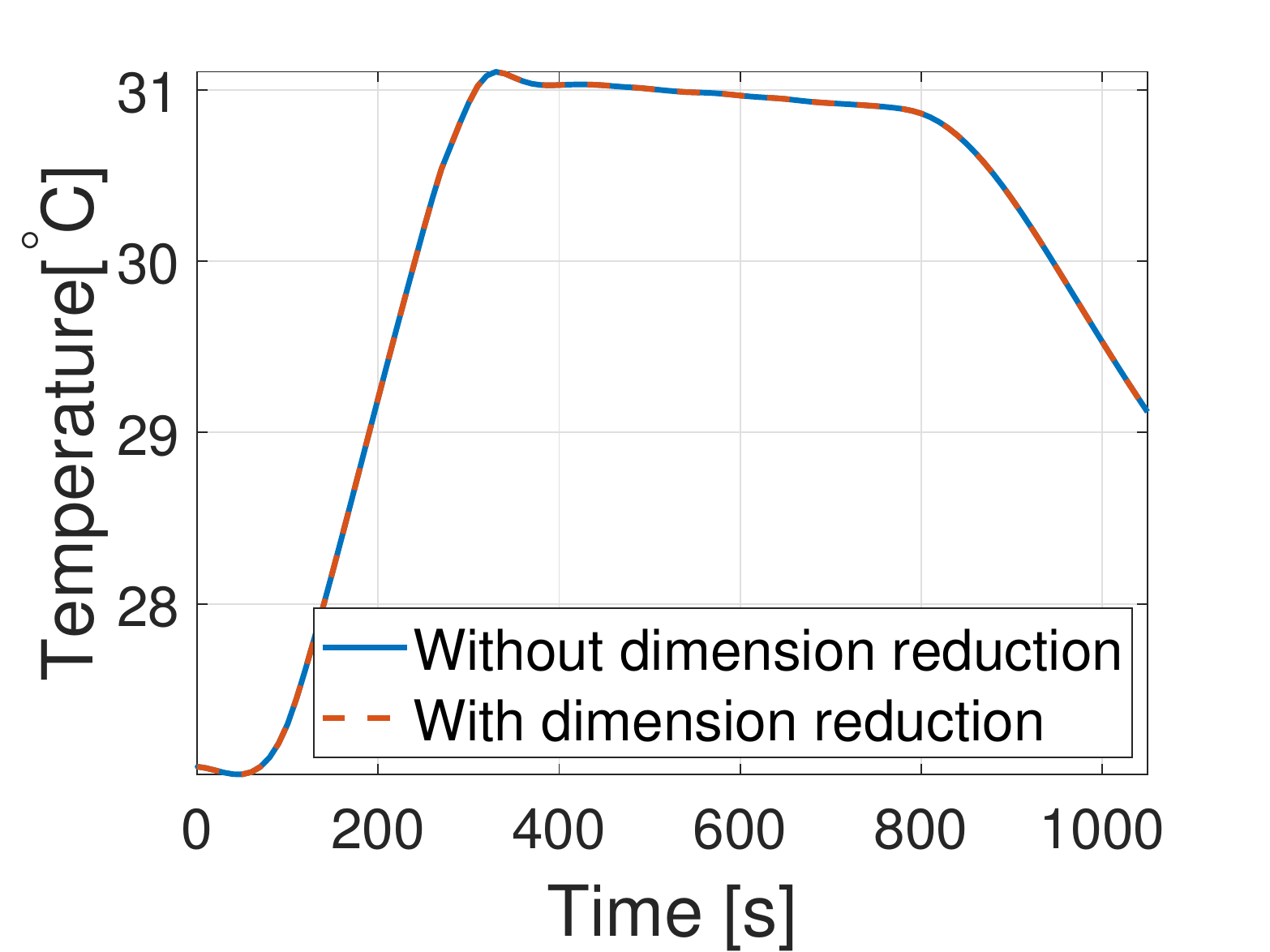}
    \caption{}
  \end{subfigure}
 
  \begin{subfigure}[c]{0.35\textwidth}
    \centering\includegraphics[width=\textwidth]{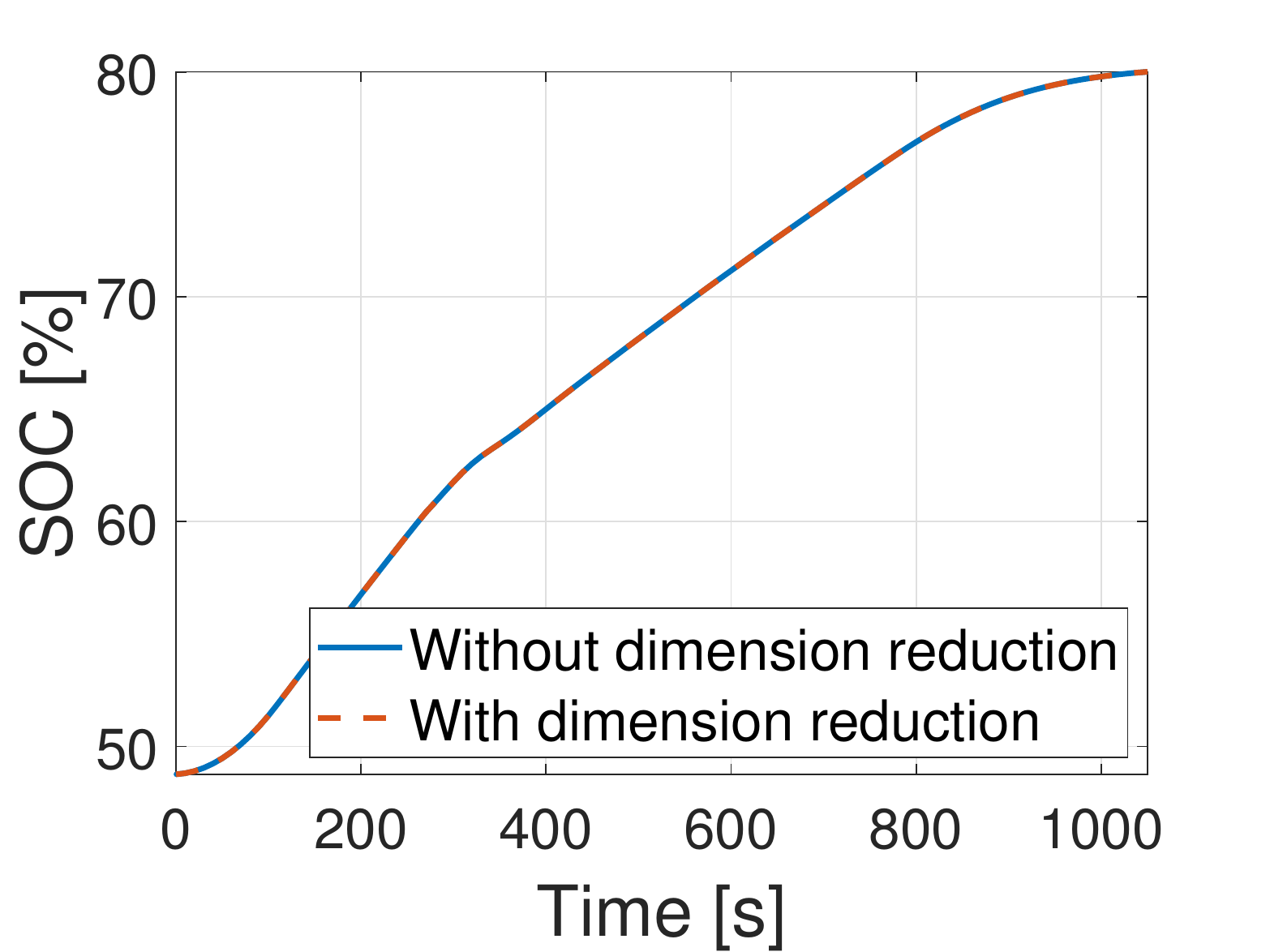}
    \caption{}
  \end{subfigure}
  \begin{subfigure}[c]{0.35\textwidth}
    \centering\includegraphics[width=\textwidth]{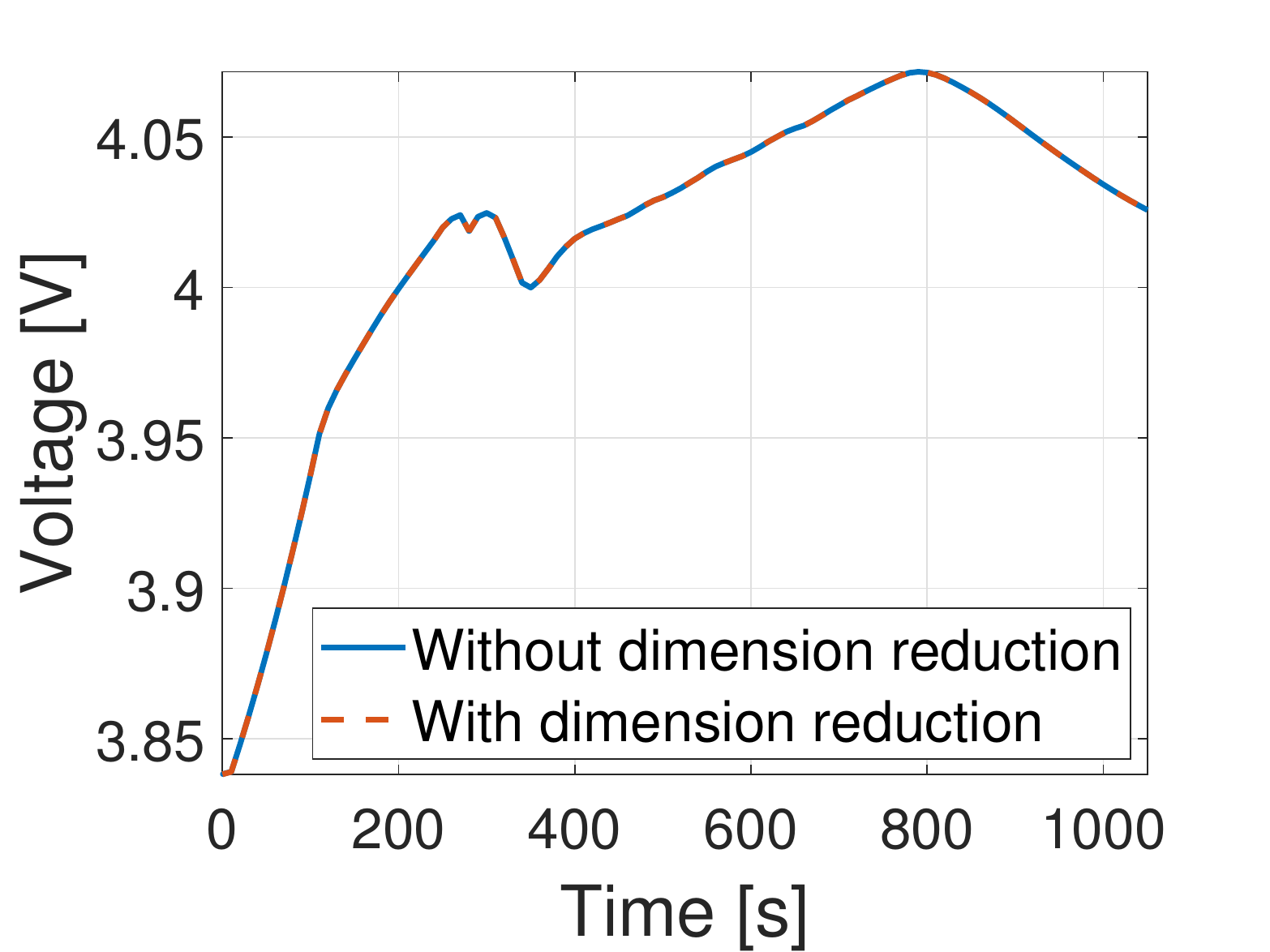}
    \caption{}
  \end{subfigure}
  \caption{\small{Simulation results of original DeePC and PCA-based DeePC}}
  \label{fig:DeePC_Comp}
\end{figure*}

\subsection{Performance of DeePC with Dimension Reduction}
The PCA-based dimension reduction method is applied to the aforementioned DeePC formulation, to demonstrate its ability of retaining optimal control performance while significantly reducing the computational complexity. Specifically, the Li-ion battery simulator is initialized under the conditions introduced in Section~\ref{subsec:setup}, and the control objective is to charge the SOC from $49\%$ to $80\%$. According to the method detailed in Section~\ref{sec:eff-DeePC}, the matrix $A$ is factorized with SVD, and a truncated right-singular matrix $V_{[1:l]}$ is chosen with $l=700$ (the original total number of columns is $l_{max}=1791$) to construct the matrix $\bar{A}$. Figure~\ref{fig:DeePC_Comp} shows the performance comparison between the original DeePC and our PCA-based efficient DeePC. The close match of system input and outputs clearly demonstrates that the performance of the proposed efficient DeePC is as effective as that of the original DeePC. Moreover, the cost and computation time of these two methods are concluded in Table~\ref{tab:comp}. The costs of both DeePC methods are basically the same, while the time-per-step optimization of the proposed efficient DeePC is significantly less. %The average optimization time of the efficient DeePC approach is only a quarter of that for the original DeePC. Meanwhile, its performance and cost are barely impacted by the optimization variables of reduced dimension.

\begin{table}[t]
\caption{\small{DeePC performance metrics based on $A$ and $\bar{A}$}}
\label{tab:comp}
\centering
\begin{tabular}{c c c}
\hline
 & $A\in\mathbb{R}^{520\times 1791}$ & $\bar{A}\in\mathbb{R}^{520\times l} \; (l=700)$  \\
\hline
Cost & 261813.25 & 261813.26 \\
%\hline
Time per Step [sec] & 2.2s & 0.5s \\
\hline
\end{tabular}
\end{table}

\begin{figure*}[t]
  \centering
  \includegraphics[width=0.7\textwidth]{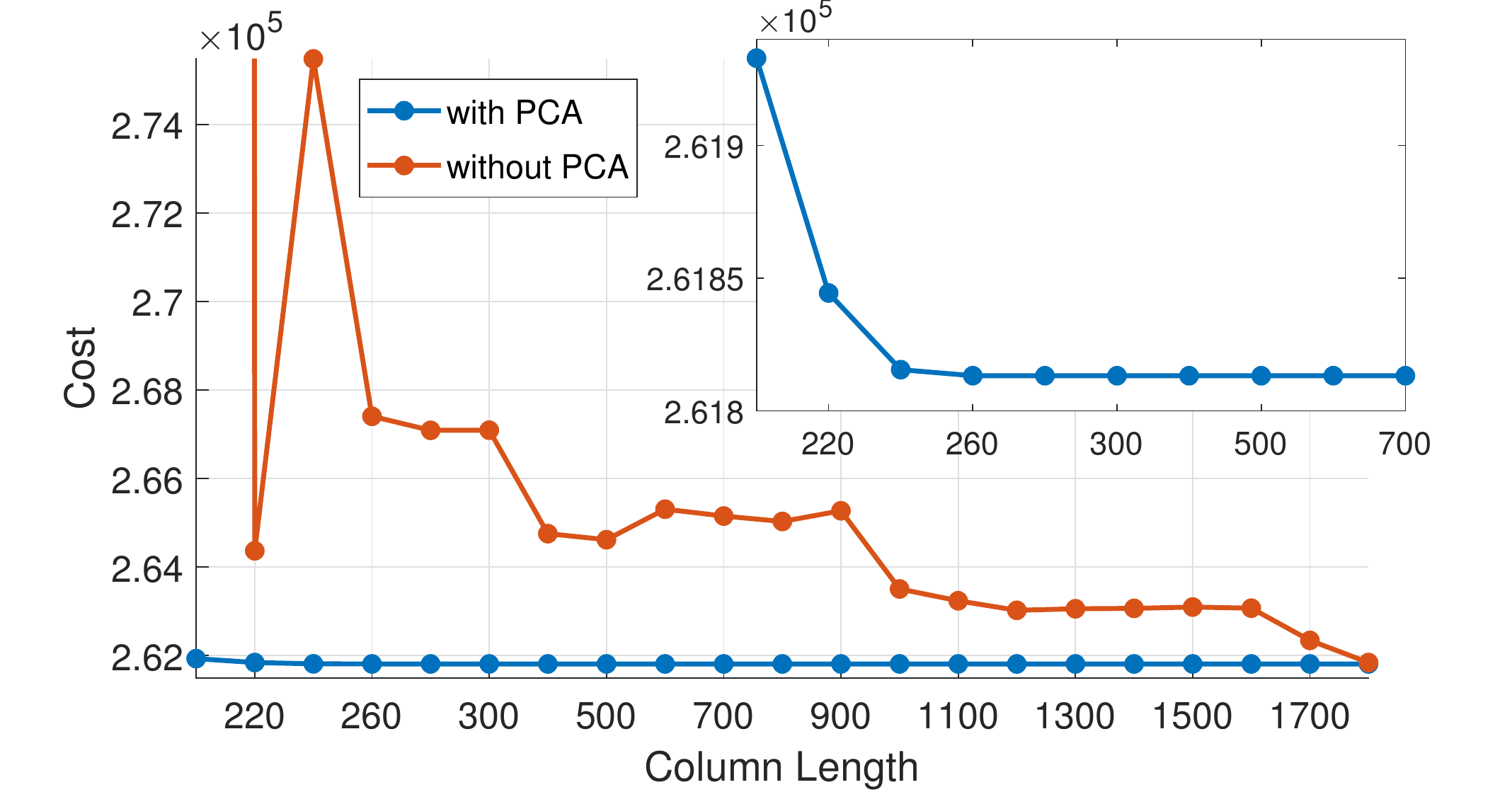}
  \caption{{\small Cost comparison between the PCA-based DeePC and original DeePC in the case of same Hankel matrix size.}}\label{fig:Pred_vs_Ori}
\end{figure*}

Further simulations are run by varying $l$ to form matrices $\bar{A}$ with different column numbers (i.e., different dimensions of the optimization variables). A comparison is conducted by directly truncating matrix $A$ to make its size exactly the same as that of $\bar{A}$ so that the computational resource used by these two cases are about the same to compare its control performance. The results are shown in Figure~\ref{fig:Pred_vs_Ori}. It can be found that for $l\in [240,1791]$, the cost function values of PCA-based DeePC can maintain stable and comparable performance within a large range. However, DeePC without PCA (i.e., direct truncation of matrix $A$) shows distinctly fluctuant cost values under different $l$. 
%According to the results shown in Figure~\ref{fig:Pred_vs_Ori}, if $l$ is chosen as 240, the computational time can be further reduced to $0.17$ second compared to the previous case with $l=700$, and the cost is 261816, with only slight increase. 
Therefore, it is clear that the PCA-based dimension reduction strategy can significantly reduce the computational complexity while retaining the optimal performance comparable to the original DeePC scheme.

\section{Conclusion}\label{sec:conclusions}
In this paper, we presented a novel DeePC approach for Li-ion battery fast charging with safety constraints. The DeePC approach only exploited input/output data for controller design, without the need for the complicated parametric battery model identification and calibration. In addition, to make DeePC online applicable with improved computational efficiency, a PCA-based dimension reduction approach was developed to significantly reduce computational time while retaining optimal system performance. Simulation results on a high-fidelity battery simulator confirmed the efficacy of the proposed scheme, with a 15\% reduction in charging time as compared to the CC-CV-$1.2C$ benchmark while satisfying all system constraints. Future work will include the integration of battery degradation metrics in the DeePC framework.



%\biboptions{sort&compress}
\bibliographystyle{elsarticle-num} 
%\bibliographystyle{plain}
%\bibliography{DeePC_ref.bib}
\bibliography{IEEEabrv,ref_abrv}


%\begin{thebibliography}{00}


%\bibitem{}

%\end{thebibliography}
\end{document}